\documentclass[twocolumn]{jpsj3}
\title{Spin-Wave Spectrum in `Single-Domain'  Magnetic Ground State\\%
of Triangular Lattice Antiferromagnet CuFeO$_2$}

\author{Taro Nakajima\thanks{E-mail address: nakajima@nsmsmac4.ph.kagu.tus.ac.jp}, 
Setsuo Mitsuda, Tendai Haku, Kohei Shibata, Keisuke Yoshitomi, Yukio Noda$^1$, Naofumi Aso$^2$, Yoshiya Uwatoko$^3$ and Noriki Terada$^4$}

\inst{Department of Physics, Faculty of Science, Tokyo University of Science, Tokyo 162-8601\\
$^{1}$Institute of Multidisciplinary Research for Advanced Materials, Tohoku University, Sendai 980-8577\\
$^2$Faculty of Science, University of the Ryukyus, Nishihara, Okinawa 903-0213\\
$^3$Institute for Solid State Physics, University of Tokyo, Kashiwa, Chiba 277-8581\\
$^4$National Institute for Materials Science, Sengen 1-2-1, Tsukuba, Ibaraki 305-0044}

\abst{By means of neutron scattering measurements, we have investigated spin-wave excitation in a collinear four-sublattice (4SL) magnetic ground state of a triangular lattice antiferromagnet CuFeO$_2$, %
which has been of recent interest as a strongly frustrated magnet, a spin-lattice coupled system and a multiferroic. %
To avoid mixing of spin-wave spectrum from magnetic domains having three different orientations reflecting trigonal symmetry of the crystal structure, %
we have applied uniaxial pressure on $[1\bar{1}0]$ direction of a single crystal CuFeO$_2$. %
By elastic neutron scattering measurements, we have found that only 10 MPa of the uniaxial pressure results in almost `single domain' state in the 4SL phase. %
We have thus performed inelastic neutron scattering measurements using the single domain sample, and have identified two distinct spin-wave branches. %
The dispersion relation of the upper spin-wave branch cannot be explained by the previous theoretical model [R. S. Fishman: J. Appl. Phys. {\bf 103} (2008) 07B109]. % 
This implies the importance of the lattice degree of freedom in the spin-wave excitation in this system, because the previous calculation neglected the effect of the spin-driven lattice distortion in the 4SL phase. %
We have also discussed relationship between the present results and the recently discovered "electromagnon" excitation. %
}

\kword{CuFeO$_{2}$, spin wave, spin frustration, inelastic neutron scattering}

\begin{document}
\maketitle

\section{Introduction}

\begin{figure}[t]
\begin{center}
	\includegraphics[clip,keepaspectratio,width=0.42\textwidth]{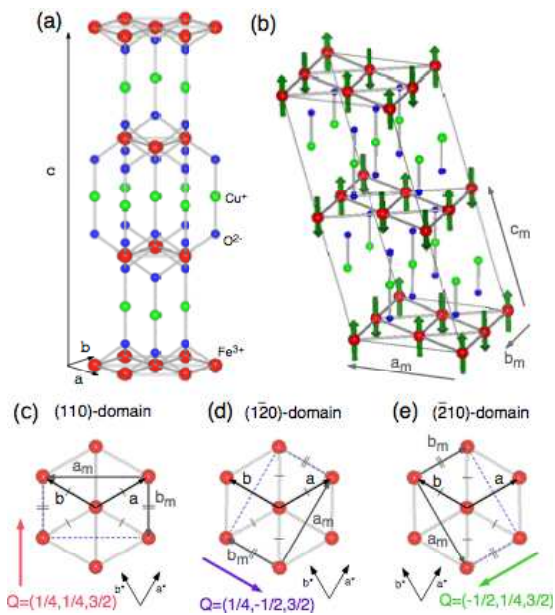}
	\caption{(Color online) (a) Crystal structure of CuFeO$_2$ with the hexagonal basis. (b) Magnetic structure of the 4SL phase with the monoclinic basis. %
	[(c)-(e)] Relationships between the hexagonal and the monoclinic bases in the (c) (110)-, (d) $(1\bar{2}0)$- and (e) $(\bar{2}10)$-domains. }
	\label{CrystStr}
\end{center}
\end{figure}

From 1990s, a delafossite compound CuFeO$_2$ has been extensively studied as a frustrated spin system.\cite{Mitsuda_1991,Mekata_1993} %
Because of triangular geometry of magnetic Fe$^{3+}$ ions and antiferromagnetic interactions between them, this system has vast degeneracy around the ground state, in which a collinear four-sublattice (4SL) antiferromagnetic order is realized.\cite{Mitsuda_1991} %
Therefore, application of magnetic fields or substitution of a small amount of nonmagnetic ions for magnetic Fe$^{3+}$ ions induces a variety of magnetic phases and unconventional magnetic phase transitions.\cite{Mitsuda_2000,Terada_x_T,Terada_CFO_Pulse,Seki_PRB_2007,Ga-induce,CFRO} % 
Moreover, the first magnetic-field induced phase from the 4SL phase has been recently focused as a spin-driven ferroelectric phase,  in which an elliptic screw-type magnetic ordering breaks inversion symmetry of the system and accounts for the charge polarized state.\cite{Kimura_CuFeO2,SpinNoncollinearlity,CompHelicity} %
To understand these various magnetic phase transitions, it is essential to establish the spin Hamiltonian of this system. %
One of the most effective way to determine the spin Hamiltonian is to measure spin-wave dispersion relation by inelastic neutron scattering measurements. %
Although several previous studies have reported inelastic neutron scattering measurements in the 4SL phase,\cite{Terada_CFO_SW1,Terada_CFO_SW2,Ye_CFO_SW} there still remains some ambiguity in determination of the spin-wave dispersion relations from the measured spectrum. %
This is due to existence of three magnetic domains reflecting trigonal symmetry of the crystal structure. %
Figures \ref{CrystStr}(a) and \ref{CrystStr}(b) show the crystal structure of CuFeO$_2$ and the 4SL magnetic structure, whose magnetic propagation wave vector is $(\frac{1}{4},\frac{1}{4},\frac{3}{2})$, respectively. %
Owing to the threefold rotational symmetry about the $c$ axis, there are three magnetic domains whose wave vectors of %
$(\frac{1}{4},\frac{1}{4},\frac{3}{2})$, $(\frac{1}{4},-\frac{1}{2},\frac{3}{2})$ and $(-\frac{1}{2},\frac{1}{4},\frac{3}{2})$ are crystallographically equivalent to each other. %
Therefore, the magnetic excitation spectrum measured in the previous studies are superpositions of the spin-wave spectrum from the three magnetic domains with different orientations.
On the other hand, recent synchrotron radiation x-ray diffraction studies have revealed that CuFeO$_2$ exhibits trigonal to monoclinic crystal structural transition associated with the paramagnetic to antiferromagnetic phase transition.\cite{Terada_CuFeO2_Xray,Ye_CuFeO2} %
This structural transition also results in three monoclinic structural domains, and there is a one-to-one correspondence between the monoclinic structural domains and the magnetic domains, as shown in Figs. \ref{CrystStr}(c)-\ref{CrystStr}(e). %
This implies the possibility that volume fractions of the three magnetic domains can be controlled by application of pressure, which directly affects lattice degree of freedom. %
In the present study, we have demonstrated that application of uniaxial pressure on a single crystal CuFeO$_2$ results in almost `single-domain' 4SL phase. %
Using the single-domain sample, we have performed inelastic neutron scattering measurements, and have identified the spin-wave dispersion relations for the 4SL phase. %

\section{Preliminary Details and Experiments}

CuFeO$_2$ exhibits two magnetically ordered phases in zero magnetic field. The high temperature phase is an incommensurate collinear magnetic phase referred to as partially disordered (PD) phase,\cite{Mitsuda_1994_PD} whose propagation wave vector is $(q,q,\frac{3}{2})$ with $q=0.202\sim 0.210$, and the low temperature phase is the 4SL phase. %
The transition temperatures for the two magnetic phases are $T_{\rm N1} = 14$ K and $T_{\rm N2} \sim 11$ K, respectively. %
Below $T_{\rm N1}$, the crystal structural transition occurs and the three magnetic domains are formed. %
In this paper, we refer to the three domains with the wave vectors of $(\frac{1}{4},\frac{1}{4},\frac{3}{2})$, $(\frac{1}{4},-\frac{1}{2},\frac{3}{2})$ and $(-\frac{1}{2},\frac{1}{4},\frac{3}{2})$ as $(110)$-, %
$(1\bar{2}0)$- and $(\bar{2}10)$-domains, respectively. %
Relationships between the monoclinic and the hexagonal bases in each domain are shown in Figs. \ref{CrystStr}(c)-\ref{CrystStr}(e). %
To distinguish between the two bases, the subscript "m" has been added  to the monoclinic notation when referring to the basis, wave vectors and reciprocal lattice indices. %
The previous x-ray diffraction studies have revealed that with decreasing temperature, $a_{\rm m}$ contracts, and on the contrary, $b_{\rm m}$ elongates in each domain.\cite{Terada_CuFeO2_Xray,Ye_CuFeO2} %
We thus anticipated that uniaxial pressure applied perpendicular to the $c$ axis favors the magnetic domains whose $a_{\rm m}$ axis lies along the pressure, and, on the contrary, suppresses those whose $b_{\rm m}$ axis lies along the pressure. %
Therefore, we cut a single crystal of CuFeO$_2$ grown by the floating zone technique\cite{Zhao_FZ} into thin plate shape ($\sim 6\times$10$\times$2 mm$^3$) with the widest surface normal to the $[1\bar{1}0]$ direction, on which we applied uniaxial pressure so as to maximize the volume fraction of the (110)-domain. %
The sample was set in a uniaxial pressure cell developed by Aso \textit{et al.}\cite{Pressurecell} %
We applied, at room temperature, 10 MPa of uniaxial pressure on the sample. %
CuBe disk springs were set in the pressure cell to keep the applied pressure. %
In order to evaluate the volume fractions of the three magnetic domains, we performed four-circle neutron diffractions measurement using FONDER installed at JRR-3 in Japan Atomic Energy Agency (JAEA). %
The incident neutron beam with wavelength 1.240 \AA \  was obtained by a Ge(311) monochromator. %
The sample with the pressure cell was mounted on a closed-cycle He-gas refrigerator, and was cooled down to 2.5 K. %

\begin{figure}[t]
\begin{center}
	\includegraphics[clip,keepaspectratio,width=0.40\textwidth]{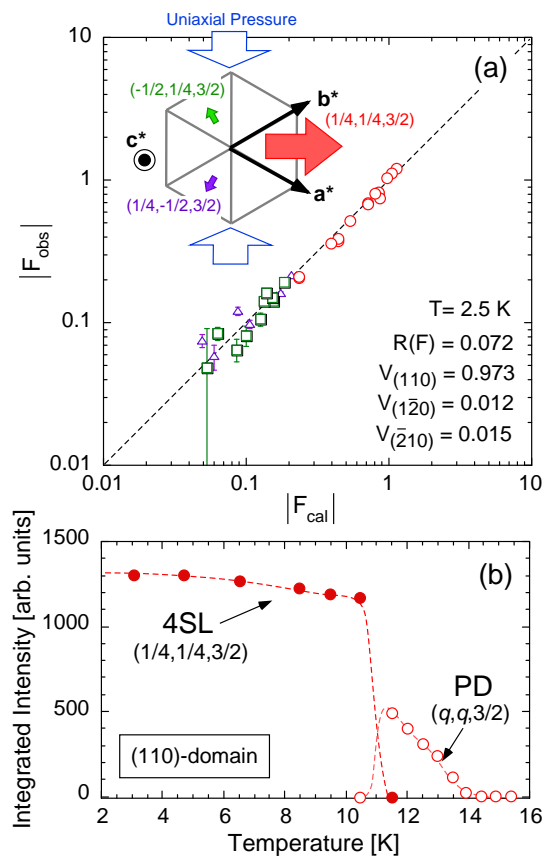}
	\caption{(Color Online) (a) Comparison between the observed and the calculated magnetic structure factors. %
	Open circles, triangles and squares denote data for  the (110)-, $(1\bar{2}0)$- and $(\bar{2}10)$-domains, respectively. %
	Inset shows schematic drawing of direction of the uniaxial pressure (open blue arrows) and orientations of the three magnetic domains. %
	Filled red, green and purple arrows denote directions of $c$-plane-projections of the magnetic modulation wave vectors; the sizes of the arrows qualitatively show the volume fractions of the three domains. 
	(b) Temperature variation of integrated intensities of magnetic reflections corresponding to the PD and the 4SL magnetic orderings in the (110)-domain. }
	\label{FONDER}
\end{center}
\end{figure}

\begin{figure*}[t]
\begin{center}
	\includegraphics[clip,keepaspectratio,width=0.94\textwidth]{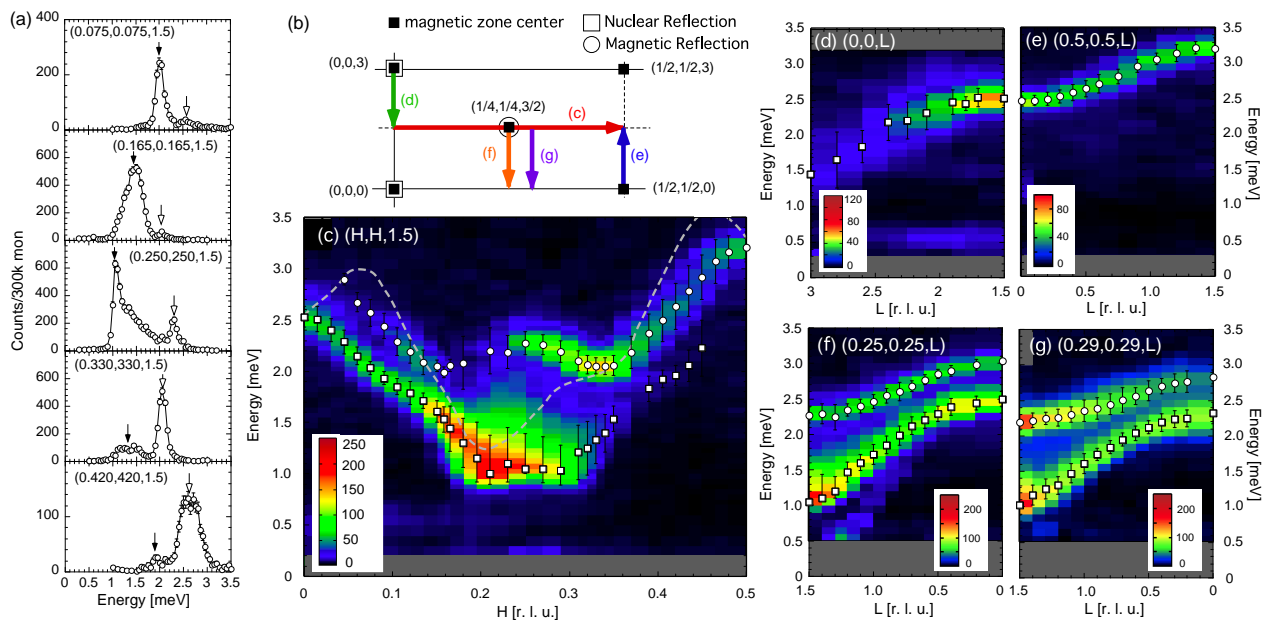}
	\caption{(Color Online) (a) Typical profiles of constant-$Q$ scans  in the present inelastic neutron scattering measurement. %
	Open and filled allows denote peak positions corresponding to the HE and the LE branches. %
	(b) Reciprocal lattice map of the $(H,H,L)$ scattering plane. Open squares and a circle denote positions of nuclear and magnetic Bragg reflections, respectively. Filled squares denote positions of magnetic zone center. %
	[(c)-(g)] Spin-wave spectrum along (c) $(H,H,\frac{3}{2})$, (d) $(0,0,L)$, (e) $(\frac{1}{2},\frac{1}{2},L)$, (f) $(\frac{1}{4},\frac{1}{4},L)$ and (g) $(0.29,0.29,L)$ lines. %
	Circles and squares represent peak positions in constant-$Q$ scans, and vertical bars represent full width at half maximum of the peaks. %
	Dashed line in (c) shows the asymmetric branch observed in the previous studies\cite{Ye_CFO_SW,Terada_CFO_SW2} (taken from Ref. \citen{Ye_CFO_SW}).}
	\label{SW}
\end{center}
\end{figure*}

After the four-circle neutron diffraction measurement, we have performed inelastic neutron scattering measurement using the identical sample in the pressure cell, which has kept the 10 MPa of uniaxial pressure throughout the two neutron scattering experiments. %
For this measurement, we used a cold neutron triple axis spectrometer HER(C1-1) installed at JRR-3 in JAEA. %
The wave number of the scattered neutrons was fixed to $k_f = 1.3246$ \AA$^{-1}$. %
A horizontal focusing analyzer was employed. %
Energy resolution at elastic position is 0.13 meV  (full width at half maximum). %
The higher-order contaminations were removed by a cooled Be-filter and a PG filter placed in front of the sample and a room-temperature Be-filter placed in front of the analyzer. %
The sample with the pressure cell was mounted in a pumped $^4$He cryostat with $(H,H,L)$ scattering plane, on which the magnetic modulation wave vector of the (110)-domain lies, and was cooled down to 1.4 K. %

\section{Results and Discussions}
\subsection{Four-circle neutron diffraction measurements}

In the present four-circle neutron diffraction measurement, we measured intensities of 32 magnetic Bragg reflections, of which 15, 7 and 10 magnetic reflections belong to the (110)-, $(1\bar{2}0)$- and $(\bar{2}10)$-domains, respectively. %
The effect of neutron absorption was corrected by the DABEX software. %
We have performed least-square fitting analysis, in which we used the established magnetic structural parameters of the 4SL phase presented in Ref. \citen{Mitsuda_1991} and refined the volume fractions of the (110)-, $(1\bar{2}0)$- and $(\bar{2}10)$-domains, $V_{(110)}, V_{(1\bar{2}0)}$ and $V_{(\bar{2}10)}$. %
The comparison between the observed magnetic structure factor $|F_{\rm obs}|$ and the calculated values $|F_{\rm cal}|$ is shown in Fig. \ref{FONDER}(a). %
The reliability factor was obtained to be 7.2 \%, indicating that all of the observed data is well explained by the 4SL magnetic structure. %
The volume fractions of the three magnetic domains are determined to be $V_{(110)}:V_{(1\bar{2}0)}:V_{(\bar{2}10)}=0.973:0.012:0.015$. % 
This shows that the uniaxial pressure applied onto the $[1\bar{1}0]$ surface successfully produces almost `single-domain' 4SL state in which the (110)-domain dominates over the $(1\bar{2}0)$- and the $(\bar{2}10)$-domains. %

Figure \ref{FONDER}(b) shows temperature variation of integrated intensities of magnetic reflections corresponding to the PD and the 4SL magnetic orderings in the (110)-domain, which were measured on cooling. %
We found that the transition temperatures and temperature variation of the order parameters are almost the same as those in zero pressure.\cite{Ye_CuFeO2} %
This indicates that the application of 10 MPa of the uniaxial pressure does not largely affect the magnetic interactions in this system, while it does the volume fractions of the three domains. %

\subsection{Inelastic neutron scattering measurements}

Figures \ref{SW}(a) and \ref{SW}(c) show typical profiles of constant wave vector ($Q$) scans and magnetic excitation spectrum along $(H,H,\frac{3}{2})$ line, respectively. %
These data have revealed that in the 4SL phase, there are two different spin-wave branches. %
In this letter, we refer to the upper and the lower branches as high-energy (HE) and low-energy (LE) branches, respectively. %
We found that both of the branches are symmetric with respect to $H=\frac{1}{4}$. %
On the other hand, in the previous studies,\cite{Ye_CFO_SW,Terada_CFO_SW2} another asymmetric branch has been observed in the $(H,H,\frac{3}{2})$ line; %
the dispersion relation of the asymmetric branch is drawn by a dashed line in Fig. \ref{SW}(c). %
Ye \textit{et al.} have suggested that the asymmetric branch belongs to the magnetic domains whose propagation wave vectors point out of the $(H,H,L)$ scattering plane.\cite{Ye_CFO_SW} %
Actually, the present measurements have demonstrated that the asymmetric branch was not observed in the `single-domain' 4SL state, confirming the previous suggestion. %

It should be mentioned that recent theoretical work by Fishman has pointed out that the spin-wave dispersion along the $(H,H,\frac{3}{2})$ line consists of two branches.\cite{Fishman_SW}
However, in the previous inelastic neutron scattering studies on this system, only the LE branch has been identified to belong to the (110)-domain.\cite{Ye_CFO_SW,comment1} %
This is because the energies of the HE branch are close to those of the asymmetric branch, and therefore, it is quite difficult to distinguish between them. %
Removing the asymmetric branch by the applied uniaxial pressure, we have, for the first time, identified the dispersion relation of the HE branch.  %
This suggests that the magnetic domain control by applied uniaxial pressure is useful to determine spin-wave dispersion relation in spin-lattice coupled complex magnets with multi-domain structures. % 

The measured dispersion relation of the HE branch has two energy dips at $H\sim 0.33$ and $\sim 0.17$. This does not agree with the previous theoretical result by Fishman,\cite{Fishman_SW} %
indicating that the magnetic interaction parameters presented in Refs. \citen{Ye_CFO_SW} and \citen{Fishman_SW} are not sufficient to explain the spin-wave dispersion relation in this system. %
We have also measured the spin-wave dispersion along the $c^*$ axis at $H=K=0.25$ and $0.29$, as shown in Figs. \ref{SW}(f) and \ref{SW}(g). %
These results have revealed that the HE branch is less dispersive as compared to the LE branch. %

We now discuss the energies of the spin waves at the magnetic zone center, specifically, $(0,0,0)$, $(0,0,3)$, $(\frac{1}{4},\frac{1}{4},\frac{3}{2})$, $(\frac{1}{2},\frac{1}{2},0)$ and so on, because these energies are relevant to optical responses including ESR modes. %
As for the LE branch, the previous work by Ye \textit{et al}.\cite{Ye_CFO_SW} have clearly showed that there are two peaks at 1.1 and 1.4 meV in a constant-$Q$ scan at $(0,0,3)$ [see Fig. 2(b) in Ref. \citen{Ye_CFO_SW}]. %
Although in the present measurements, we could not identify the two-peak structure in the constant-$Q$ scan at (0,0,3) due to lack of intensity, the asymmetric peak profile in the constant-$Q$ scan at $(\frac{1}{4},\frac{1}{4},\frac{3}{2})$ suggests that the LE branch is split into two branches at the magnetic zone center [see Fig. \ref{SW}(a)]. %
These energies show good agreements with the antiferromagnetic resonance modes around 270 and 340 GHz observed in the previous ESR measurements on CuFeO$_2$.\cite{Fukuda_CFO_ESR,Kimura_CFO_ESR} %

As for the HE branch, we have measured the spin-wave dispersion along the $(\frac{1}{2},\frac{1}{2},L)$ line as shown in Fig. \ref{SW}(e), and have found that the HE branch has an energy of $2.47$ meV at $(\frac{1}{2},\frac{1}{2}, 0)$. %
This energy  agrees with the ESR signal around 600 GHz observed in the previous measurement.\cite{Kimura_CFO_ESR} %
However, in Ref. \citen{Kimura_CFO_ESR}, Kimura \textit{et al}. have argued that their conventional spin-wave calculation cannot reproduce the ESR signal around 600 GHz. %
On the other hand, quite recently, Seki \textit{et al.} have performed terahertz time-domain spectroscopy in the 4SL phase. They have observed an `electromagnon' (electric-field-active magnon) excitation around 2.3 meV,\cite{Seki_electromagnon} which agrees with the energy of the HE branch at $(\frac{1}{4},\frac{1}{4},\frac{3}{2})$. %
These results suggest that the HE branch has novel spin dynamics coupled with lattice or charge degree of freedom. %
It should be also mentioned that the reason for the finite difference between the energy of the HE branch at $(\frac{1}{2},\frac{1}{2}, 0)$ $(2.47$ meV) and that at $(\frac{1}{4},\frac{1}{4},\frac{3}{2})$ $(2.30$ meV) is not clear at this moment. %

\section{Conclusion}

In conclusion, we have investigated spin-wave excitation in the 4SL magnetic ground state of CuFeO$_2$. %
To avoid mixing of the spin-wave spectrum from the three magnetic domains, we applied 10 MPa of uniaxial pressure onto the $[1\bar{1}0]$ surface of the single crystal CuFeO$_2$. %
The present four-circle neutron diffraction measurements have demonstrated that the application of the pressure successfully produces `single-domain' 4SL phase. %
Using the identical sample, we have performed the inelastic neutron scattering measurements. %
As a result, we have successfully identified the two (LE and HE) spin-wave branches, both of which are symmetric with respect to $(\frac{1}{4},\frac{1}{4},\frac{3}{2})$. %
It should be emphasized that the dispersion relation of the HE branch does not agree with the previous theoretical work\cite{Fishman_SW}, %
which is based on the assumption that the spin-driven crystal lattice distortion\cite{Terada_CuFeO2_Xray,Ye_CuFeO2} hardly affects the spin-wave dispersion relations in the 4SL phase. %
We have thus suggested that the spin-lattice coupling is essential to understand the spin-wave excitation in this system. 
The asymmetric branch observed in the previous works\cite{Ye_CFO_SW,Terada_CFO_SW1,Terada_CFO_SW2} was not observed in the single domain 4SL state. %
This confirms the previous suggestion that the asymmetric branch arises from the magnetic domains with the magnetic modulation wave vectors pointing out of scattering plane.\cite{Ye_CFO_SW} %
We have also discussed the relationship between the energies of the spin waves at the magnetic zone center and the results of the optical measurements on this system,\cite{Fukuda_CFO_ESR,Kimura_CFO_ESR,Seki_electromagnon} suggesting that the HE branch corresponds to the "electromagnon" excitation, in which dynamical coupling between spin and dielectric polarization is realized. %

\section*{Acknowledgment}
We are grateful to Shojiro Kimura for fruitful discussions. %
The neutron scattering measurements at JRR-3 were carried out along the proposals No. 10783B and No. 10586, and were partly supported by ISSP of the University %
of Tokyo. %
This work was partly supported by Grants-in-Aid for Scientific Research gYoung Scientists (B), Grants No. 20740209h. We thank M. Yokoyama for technical support in the neutron scattering experiment with HER(C1-1). %
The images of the crystal and magnetic structures in this letter were depicted using software VESTA.\cite{VESTA} %

\end{document}